\documentclass[aps,prl,twocolumn,showpacs]{revtex4-1}
\usepackage{graphicx} 
\usepackage{amssymb}
\usepackage{float}
\begin{document}
\title
{\bf Unconventional Pairings and Radial Line Nodes in Inversion Symmetry Broken Superconductors}
\author{T. Hakio\u{g}lu$^{\bf (1,3)}$ and Mehmet G\"{u}nay$^{\bf (2,3)}$}
\affiliation{ 
${\bf (1)}$ {Program on Quantum Technologies in Energy, Energy Institute, \.{I}stanbul Technical University, 34469, \.{I}stanbul, Turkey}
\break
${\bf (2)}$ {Department of Physics, Bilkent University, 06800 Ankara, Turkey}
\break
${\bf (3)}$ {Institute of Theoretical and Applied Physics (ITAP), 48740 Turun\c{c}, Mu\u{g}la, Turkey}
}
\begin{abstract}
Noncentrosymmetric superconductors (NCSs) with broken inversion symmetry can have spin-dependent order parameters (OPs) with mixed parity which can also have nodes in the pair potential as well as the energy spectra. These nodes are distinct features that are not present in conventional superconductors. They appear as points or lines in the momentum space where the latter can have angular or radial geometries dictated by the dimensionality, the lattice structure and the pairing interaction.  

In this work we study the nodes in time reversal symmetry (TRS) preserving NCSs at the OP, the pair potential, and the energy spectrum levels. Nodes are examined by using spin independent pairing interactions respecting the rotational $C_{\infty v}$ symmetry in the presence of spin-orbit coupling (SOC). The pairing symmetries and the nodal topology are affected by the relative strength of the pairing channels which is studied for the mixed singlet-triplet, pure singlet, and pure triplet. Complementary to the angular line nodes widely present in the literature, the $C_{\infty v}$ symmetry here allows radial line nodes (RLNs) due to the nonlinear momentum dependence in the OPs. The topology of the RLNs in the mixed case shows a distinctly different characterization than the half-spin quantum vortex at the Dirac point. We apply this NCS physics to the inversion symmetry broken exciton condensates (ECs) in double quantum wells where the point and the RLNs can be found. On the other hand, for a pure triplet condensate, two fully gapped and topologically distinct regimes exist, separated by a QSHI-like zero energy superconducting state with even number of Majorana modes. We also remark on how the point and the RLNs can be manipulated, enabling an external control on the topology.
\vskip0.5truecm
corresponding author: T. Hakio\u{g}lu, fax: + 90 212 285 3885, tel: + 90 212 285 3884, e-mail: hakioglu@itu.edu.tr 
\vskip0.5truecm    
\end{abstract}
\keywords{Non-centrosymmetric Superconductivity, Topological Superconductivity, Spin-Orbit Coupling}
\pacs{71.35.-y,71.70.Ej,03.75.Hh,03.75.Mn}
\maketitle
Pairing symmetries beyond the conventional BCS have been first addressed in the B and the A phases of $^3He$\cite{He3_th,He3_exp}. Unconventional pairing states were then reported in heavy fermion\cite{heavyfermion} and the high-$T_c$ superconductors\cite{high_Tc}. It is now settled that, the inversion symmetry (IS), the time reversal , the particle-hole ($\Lambda$) and the fermion exchange ($F_X$) i.e. Pauli exclusion symmetries play fundamental role in unconventional superconducting pairing. 

In the NCSs the IS is broken. They comprise a subset of a larger class, i.e. unconventional superconductors. The broken IS is usually connected to the presence of a SOC which requires mixed parity OPs, i.e. the even parity singlet (s) is mixed with the odd parity triplet (t). The broken IS does not mean a strong triplet, but a weakly broken IS means a singlet dominant mixed state. For instance, NMR measurements yield that $Li_2Pt_3B$ is a mixed s-t state with a strong SOC\cite{PRL98_047002} whereas $Li_2Pd_3B$ is believed to be s-dominated with a weak SOC\cite{LiPdB}. On the other hand, $BaPtSi_3$\cite{PRB80_064504} as well as $SrPtAs$\cite{PRB89_140504} are known to break IS but they were reported as BCS like pure singlets. Usually, it is experimentally hard to separately identify a dominating singlet (triplet) within a mixed state from a pure singlet (triplet). 

A comprehensive understanding of the pairing mechanisms in NCS is currently far from complete\cite{Bauer_Sigrist}. The IS breaking is fundamentally important for spin dependent mixed parity OPs, but it needs to be sufficiently large for the nodes to appear. In TRS manifested NCSs nodes appear either at the time-reversal-invariant points or lines at certain angular orientations dictated by the crystal symmetry. Another crucial point is that, nodes in the OPs do not necessarily mean nodes in the pair potential or the energy spectrum. In centrosymmetric materials with tetragonal symmetry, strong Hubbard-like electronic correlations or spin fluctuations around AFM nesting can lead to the natural separation of the s and t pairing channels without an explicit need of an IS breaking\cite{CS_NCS}. On the other hand, phonon mechanisms were suggested for some NCSs\cite{Savrasov_Sarma}. Independently from the details of the mechanism, it is crucial that the interaction symmetries should allow the simultaneous presence of a sufficiently large triplet with or without a singlet. The triplet/singlet ratio as a function of momentum is therefore an important parameter in understanding the nodes. Nodes are also closely connected with the topology of the momentum space. All these factors outlined here point at the need for more simplistic approaches stressing the self-consistent handling of interactions with realistic momentum dependence as the key for a broader understanding of the physics of NCSs.

In this work we focus on four questions that can help our understanding: a) In NCSs, can we identify factors affecting the unconventional pairings without resorting to any lattice or other material dependent symmetries and interactions?, b) How does a pairing interaction affect the nodal structure of the OPs, the pair potential and the spectrum? c) Can the nodes, and hence the topology, be controlled externally? d) How does the nodal topology in the pair potential or spectrum in an NCS relate to a topological superconductor (TSC)? 

To answer these questions we use a material independent model with maximal rotational symmetry. We also confine our attention to two dimensions. The model consists of an IS breaking SOC and an isotropic, {\it spin independent} pairing interaction ${\cal V}(q)$ with repulsive and attractive parts. This {\it minimal model} has $C_{\infty v}$ as the simplest rotational symmetry with no referral to any specific discrete point group. Our conclusions are therefore expected to be applicable to the material independent and general aspects of pairing in TRS manifested NCSs such as those under weak anisotropy. With these inputs, we examine the relation between the pairing interaction, the pairing symmetries and the nodes. 

The two dimensional  mean field Hamiltonian we consider is described in the electronic basis $\Psi_{\bf k}^\dagger=(\hat{e}_{{\bf k} \uparrow}^\dagger~\hat{e}_{{\bf k} \downarrow}^\dagger~\hat{e}_{-{\bf k} \uparrow}~\hat{e}_{-{\bf k} \downarrow})$ where Nambu and the spin sectors are denoted respectively by the Pauli matrices $\tau=\{\tau_x,\tau_y,\tau_z\}$ and $\sigma=\{\sigma_x,\sigma_y,\sigma_z\}$. The Hamiltonian is\cite{basis} 
\begin{eqnarray}
{\cal H}=\sum_{\bf k} \Psi_{\bf k}^\dagger {\cal H}_{\bf k} \Psi_{\bf k}\,, \quad  {\cal H}_{\bf k}={\cal H}_{\bf k}^0+{\cal H}_{\bf k}^{soc}+{\cal H}_{\bf k}^\Delta~. 
\label{hamilt_1}
\end{eqnarray}
Here, ${\cal H}_{\bf k}^0=\tau_z \otimes \,{\widehat \xi}_{k}$ where ${\widehat \xi}_{k}=[\hbar^2 k^2/(2m)-\mu]\sigma_0 +{\widehat \Sigma}_k$, $m$ is the band mass, $\mu$ is the Fermi energy, ${\widehat \Sigma}_k$ is the $2\times 2$ self energy matrix in the spinor basis, ${\cal H}_{\bf k}^{soc}= [(S_{\bf k}\hat{e}_{{\bf k} \uparrow}^\dagger \hat{e}_{{\bf k} \downarrow}+S^*_{\bf k} \hat{e}_{{- \bf k} \uparrow} \hat{e}^\dagger_{-{\bf k} \downarrow}) +h.c]$  is the SOC Hamiltonian and $S_{\bf k}= \alpha \,k\, exp(i\phi_{\bf k})$ is the SOC. Here $k=\vert k_x+ik_y\vert$ is the inplane wavevector, $\alpha=\gamma_0 E_z $ with $\gamma_0$ is a material dependent constant \cite{winkler, BIA}, $E_z$ is an external electric field and $exp(i\phi_{\bf k})=(k_x+ik_y)/k$ is the SOC phase. The third term in Eq.(\ref{hamilt_1}) is the pairing Hamiltonian ${\cal H}_{\bf k}^\Delta=\tau_+\otimes \widehat{\Delta}_k+h.c.$ where $\tau_\pm=\tau_x\pm i\tau_y$ and $\widehat{\Delta}_k=i[\psi_{\bf k}\sigma_0+{\bf d}_{\bf k}.{\bf \sigma}]\sigma_y$ is the spin dependent mixed OP with $\psi_{\bf k}$ and ${\bf d}_{\bf k}=\{d_{x {\bf k}},d_{y {\bf k}},d_{z {\bf k}}\}$ as the even singlet ($\psi_{\bf k}=\psi_{\bf -k}=\psi_k$) and the odd triplet (${\bf d}_{\bf k}=-{\bf d}_{-\bf k}$) respectively.\cite{He3_th,Bauer_Sigrist} The mixed OP can also be written as 
\begin{eqnarray} 
\widehat{\Delta}_k=\pmatrix{\Delta_{\uparrow\uparrow}({\bf k}) & \Delta_{\uparrow\downarrow}({\bf k}) \cr 
 \Delta_{\downarrow\uparrow}({\bf k}) & \Delta_{\downarrow\downarrow}({\bf k})}
\label{full_OP}
\end{eqnarray} 
In the triplet, $d_{x{\bf k}}=(\Delta_{\downarrow\downarrow}-\Delta_{\uparrow\uparrow})/2$, $d_{y{\bf k}}=(\Delta_{\downarrow\downarrow}+\Delta_{\uparrow\uparrow})/(2i)$ are the equal-spin pairings (ESP), $d_{z {\bf k}}=(\Delta_{\uparrow\downarrow}+\Delta_{\downarrow\uparrow})/2$ is the opposite-spin-paired (OSP) triplet, whereas $\psi_k=(\Delta_{\uparrow\downarrow}-\Delta_{\downarrow\uparrow})/2$ is the singlet. Denoting the time reversal  transformation by $\Theta$, TRS is manifested when $\Delta_{\sigma \sigma^\prime}({\bf k})=\Theta: \Delta_{\sigma \sigma^\prime}({\bf k})=\lambda_{\sigma} \lambda_{\sigma^\prime}\Delta_{\bar{\sigma} \bar{\sigma}^\prime}^*({\bf -k})$ where $\lambda_{\uparrow}=1$ , $\lambda_{\downarrow}=-1$ and $\bar{\sigma}$ is anti-parallel to ${\sigma}$. When $F_X$ and TRS are simultaneously manifested, the OPs satisfy a strong condition $\Delta_{\sigma \bar{\sigma} }(\textbf{k})=\Delta^{\ast}_{\sigma \bar{\sigma} }(\textbf{k})$  implying that $\psi_{k}$ and $d_{z {\bf k}}$ are real. Additionally, the $C_{\infty v}$ symmetry requires that the order parameters are functions of $k$ only. These conditions together imply that $\psi_{k} d_{z {\bf k}}\propto (\vert\Delta_{\uparrow\downarrow}\vert^2-\vert\Delta_{\downarrow\uparrow}\vert^2)=0$. Hence  the simultaneous admixture of the singlet and the OSP triplet should be suppressed in the TRS manifested and weakly anisotropic NCSs \cite{psi_d_z}. 
   
Under these conditions all relevant pairings allowed in the ground state of ${\cal H}$ in Eq.(\ref{hamilt_1}) are listed in Table.\ref{TRS_cases} as i) a mixed singlet-ESP triplet ({\it s-t}$_{ESP}$) in TRS and spontaneously broken TRS (SBTRS) phases, ii) a pure {\it s} in TRS phase, and iii) two pure triplet ({\it t}$_{ESP}$) and ({\it t}$_{OSP}$) respectively in TRS and SBTRS phases. 

\begin{table}
\begin{tabular}{l|l|l|l|l|l}
\hline
Case & TRS & IS & ${ \Delta}_{\sigma \sigma}({\bf k})$\,\,(ESP) & $d_{z {\bf k}}$ (OSP) & $\psi_{k}$ (OSP)\\
\hline 
i & $\checkmark$ & $\checkmark$ & $0$ & $0$ & $\psi_{k}$ (real) \\
\hline
ii & $\checkmark$  & $\times$ & $\lambda_{\sigma} F_{k} e^{i \lambda_{\sigma} \phi_{\bf k}}$ & $0$ & $\psi_{k}$ (real) \\
iii & $\checkmark$  & $\times$ & $ 0$ & $0$ & $\psi_{k}$ (real) \\ 
iv & $\checkmark$  & $\times$ & $\lambda_{\sigma} F_{k} e^{i \lambda_{\sigma} \phi_{\bf k}}$ & $0$ & $0$  \\
\hline
v & $\times$  & $\times $ & $0$ & $ D_k e^{\pm i \phi_{\bf k}}$ & $0$ \\
vi & $\times$  & $\times$ & $\lambda_{\sigma} F_{k} e^{i \lambda_{\sigma} \phi_{\bf k}} e^{i \theta^{(t)}_{k}}$ & $0$ & $\psi_{k} e^{i \theta^{(s)}_{k}}$ \\
\hline
\end{tabular} 
\caption{Possible configurations allowed in the minimal model with $C_{\infty v}$ symmetry for the s-t pairing. Here $\sigma=(\uparrow,\downarrow)$ and we consider manifested/broken TRS and IS. Here $\psi_k$, $F_k$ and $D_k$ are radial functions of $k$. Note that the cases i-vi are allowed in the minimal model irrespective of the isotropic and spin-independent pairing interaction $V(q)$.}
\label{TRS_cases}
\end{table} 

In the TRS phase, the triplet is dictated by unitarity to have the form ${\bf d}_{\bf k}=(-F_k \cos\phi_{\bf k},F_k \sin\phi_{\bf k},0) $ where $F_k$ is the ESP strength. In NCSs, the TRS preserving {\it m}-state is experimentally the most common ground state, with $Li_2Pt_3B$\cite{PRL98_047002}, $CePt_3Si$\cite{PRL94_197002} and $ CaTSi_3 $ (T:Ir,Pt)\cite{PRB90_104504} as few examples. As far as the phases in the minimal model are concerned, the {\it m}-state as energetically the most stable configuration in almost all parameter ranges of the pairing interactions used, unless one of the angular momentum channels is specifically turned off. The TRS preserving pure ${\it t}_{ESP}$ is similar to the $^3He$-B phase (BW state). In the SBTRS phase a ${\it t}_{OSP}$ is found similar to the $^3He$-A phase (ABM state, case v). The other SBTRS solution is a mixed state like in $LaNiC_2$\cite{LaNiC} (case vi). Hence, the minimal model alone, characterized by the $C_{\infty v}$ symmetry, is capable of producing a number of common pairing symmetries respecting or violating the TRS as shown in Table.\ref{TRS_cases}. In this work, we will confine ourselves only to the TRS regime described by the cases i-iv in the Table.\ref{TRS_cases}.    
\begin{figure}
    \includegraphics [scale=0.3]{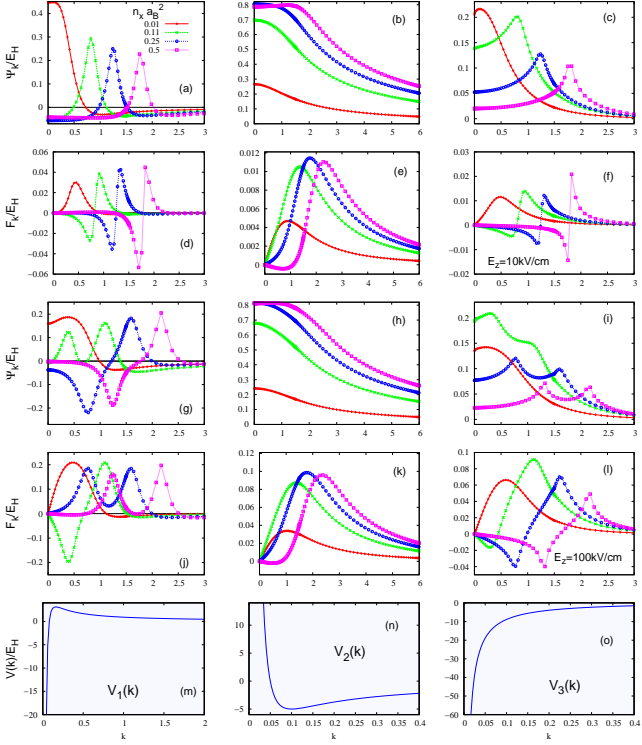}
  \label{FP}\caption{(Color online) Mixed singlet $\psi_k$ and ESP triplet $F_k$ solutions of Eq.(\ref{hamilt_1}) as a function of $k$ for the pairing potentials $V_{1,2,3}(q)$ (m-o)in the TRS phase at different SOCs for $E_z=10 kV/cm$ (a-f), and $100 kV/cm$ (g-l) and for average density of particles $\bar{n}_x=n_xa_B^2=0.01, 0.11, 0.25, 0.4$ shown [case-ii in Table.(\ref{TRS_cases})]. For comparison between the results, all energies and lengths in all figures are scaled by the Hartree energy $E_H\simeq 12 meV$ and the exciton Bohr radius $a_B\simeq 100 \AA$. Each column of figures represent the numerical solutions using the pairing potentials described in the bottom of that column.}
\label{mixed_OPs_bands}
\end{figure}  
The mean field calculations yield that the s-t OPs are coupled in the minimal model by,   
\begin{eqnarray}
\psi_k=-\frac{1}{A}\,\sum_{ k^\prime,\lambda}{\cal V}_s(k, k^\prime)\, \frac{\tilde{\Delta}_{k^{\prime}}^{\lambda}}{4E_{k^{\prime}}^{\lambda}} \Bigl\{f(E_{k^{\prime}}^{\lambda})- f(-E_{k^{\prime}}^{\lambda})\Bigr\} \nonumber \\
\label{coupled_OPs} \\ 
F_k=\frac{1}{A}\,\sum_{k^\prime,\lambda}{\cal V}_t(k, k^\prime)\, \frac{\lambda \tilde{\Delta}_{k^{\prime}}^{\lambda}}{4E_{k^{\prime}}^{\lambda}} \Bigl\{f(E_{k^{\prime}}^{\lambda})- f(-E_{k^{\prime}}^{\lambda})\Bigr\}
 \nonumber
 \end{eqnarray} 

where $\tilde{\Delta}_k^{\lambda}=(\psi_k -\lambda \gamma_k F_k)$ with the $\lambda=\pm$ signs refer to the SOC dependent splitting, $\gamma_{k}=sgn(|G_{\bf k}|{\xi}_{k}-F_{k}\psi_{k})$ with $G_{\bf k}=S_{\bf k}+(\Sigma_k)_{\uparrow\downarrow}$. Here $(\Sigma_k)_{\uparrow\downarrow}$ is the nondiagonal element of the self-energy matrix as given similarly to Eq.(\ref{full_OP}) and $ f(x)=1/[exp(\beta x)+1] $ is the Fermi-Dirac factor. The eigen energies are 
\begin{eqnarray}
E^{\lambda}_k= \sqrt{(\tilde{\xi}_k^{\lambda})^2+(\tilde{\Delta}_k^{\lambda})^2} 
\label{e_values_m}
\end{eqnarray}
 where $\tilde{\xi}_k^{\lambda}= {\xi}_{k}+\lambda \gamma_{k}  |G_{\bf k}| $. In NCS, the presence of SOC naturally separates the s and t pairing channels as $V_s$ and $V_t$ in Eq.(\ref{coupled_OPs}), and a spin-dependent interaction, like Hubbard's $U$ is not essentially needed for the s-t channel separation\cite{Manske}. The pairing interaction $V(q)$ is isotropic and spin independent with the angular momentum expansion $V(q)=\sum_{n=-\infty}^{\infty} \tilde{\cal V}_n(k,k^\prime)e^{in\phi_{{\bf k}{\bf k}^\prime}}$ where ${\bf q}={\bf k}-{\bf k}^\prime$, $\phi_{{\bf k}{\bf k}^\prime}=(\phi_{\bf k}-\phi_{{\bf k}^\prime})$ and $n$ is the angular momentum quantum number. The s-t channel separation in Eq's.(\ref{coupled_OPs}) is specifically given by\cite{phonons} 
\begin{eqnarray}    
V_s(k,k^\prime)&=&\langle V(q)\rangle_a=\tilde{\cal V}_0(k,k^\prime) \nonumber \\
\label{singlet_triplet_int}\\
V_t(k,k^\prime)&=&\langle V(q) \cos\phi_{{\bf k}{\bf k^\prime}}\rangle_a={\it Re}\{\tilde{\cal V}_1(k,k^\prime)\} \nonumber
\end{eqnarray}
with $\langle...\rangle_a$ describing the angular average over the spin-orbit phase $\phi_{{\bf k}{\bf k}^\prime}$. 
   
The model interactions $V(q)$ we use here have attractive and repulsive parts in the short/long wavelength limits. The collective excitations such as spin/charge fluctuations and phonons comprise the attractive part of $V(q)$ whereas its repulsive part is dominantly Coulombic. In order to investigate the nodes in focus of our first question, we consider three different and complementary types which are a) attractive in long wavelengths with a repulsive tail in shorter wavelengths as $V_1(q)=-A/q^2+B/q$ with $A,B >0$, b) repulsive in long and attractive in shorter wavelengths as the opposite of the first case, i.e. $A,B <0$ for $V_2(q)$. The third model is motivated by the EC in double quantum wells where the pairing is attractive and Coulombic as c) $V_3(q)=-e^2$ $exp(-qD)/(2\epsilon q)$ with $D$ describing the double quantum well separation. The EC in bulk or structural IS broken semiconductors is a promising laboratory to examine the unconventional pairing in NCS\cite{TH_PRLs,EC_vs_NCS}. Recently EC has also drawn attention in connection with the TSCs\cite{Nakosai-Wang}.    
\begin{figure*}
    \includegraphics [width=\textwidth]{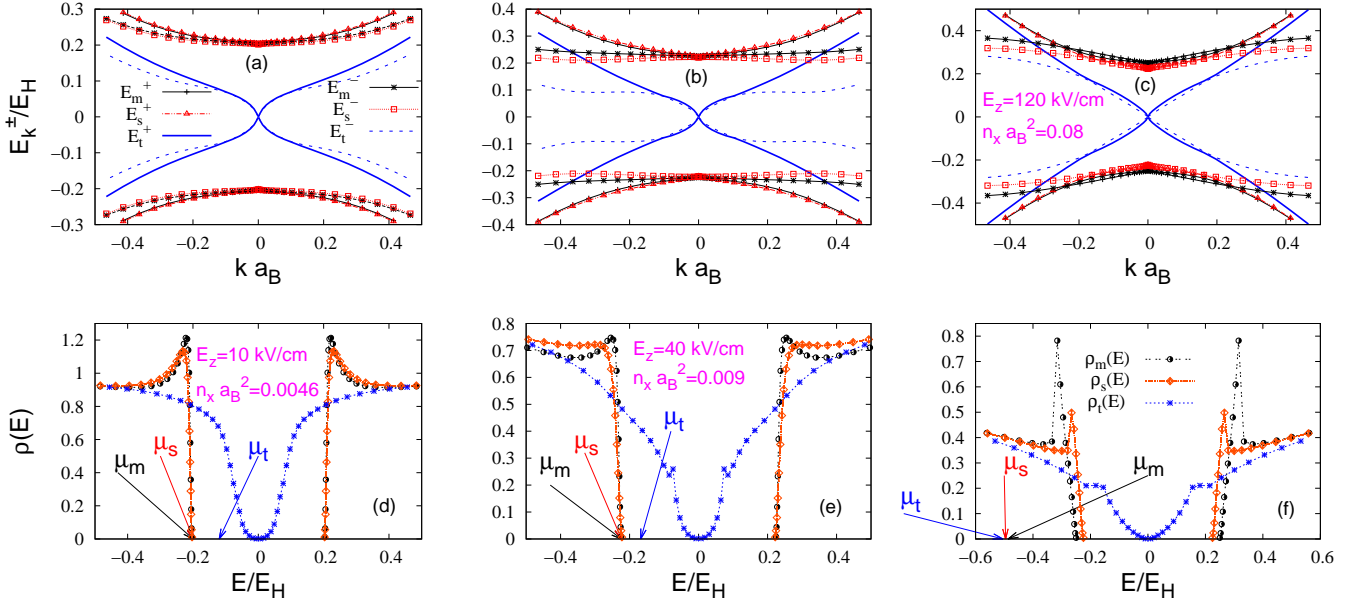} 
\vskip-0.4truecm
  \label{figur}\caption{(Color online) Mixed(${\it s-t}_{ESP}$), pure (${\it s}$) and the pure (${\it t}_{ESP}$) solutions are compared in their energy bands and DOS. The color coding in (a) and (d) apply to all figures, whereas $E_z$ and $\bar{n}_x$ values apply to vertically separated plots.}
\label{DOS}
\end{figure*} 
The numerical solutions of Eq's.(\ref{coupled_OPs}) at zero temperature are shown in Fig.\ref{mixed_OPs_bands} for $V_1(q)$ in (a,d,g,j), $V_2(q)$ in (b,e,h,k) and $V_3(q)$ in (c,f,i,k). Our observation is that, the nodes in the triplet OPs as well as the triplet/singlet (t/s) ratio are enhanced by the attractive singularities in the interaction.  
In these solutions, a rich nodal structure is revealed for $V_1(q)$ and, a large t/s ratio is obtained by increasing the SOC. In $V_2(q)$ however, and in contrast to $V_1(q)$, the attractive part is extended in a large $q$ region and there is no singularity. As a result, a weak t/s ratio is obtained with no significant nodal structure.  

We now turn our attention to $V_2(q)$. With an almost constant attractive part in intermediate $q$ regions, this interaction is like a sum of a repulsive Coulomb and a weak BCS type electron-phonon interactions. The weak momentum dependence in this BCS-like part is responsible for the poor t/s ratio in Fig.\ref{mixed_OPs_bands}(b,e,h,k). On the other hand, a phonon mediated attractive interaction can be strongly momentum dependent and can lead to a strong triplet pairing. Recently, an IS-breaking acoustic phonon mediated interaction was considered for $Bi_2Se_3$\cite{Savrasov_Sarma}. There, the pairing interaction is supported by a strong singularity at $q=0$ which overcomes the screened Coulomb repulsion in the same range, producing an effective interaction similar to $V_1(q)$ with a large t/s ratio. Finally, Fig.\ref{mixed_OPs_bands}.(c-f-i-l) indicates that the solution for $V_3(q)$ has a similar nodal structure to that of $V_1(q)$.  

These three interaction potentials above can have their origin in completely different mechanisms. A comparison of the solutions for $V_{1,2,3}(q)$ reveals that, whatever the driving mechanism is, the triplet nodes are enhanced if the potential has a strongly attractive part in the long wavelengths. This intricate connection between the momentum dependence and the nodes is also a signature justifying the need for an exact numerical solution of Eq's.(\ref{coupled_OPs}). 

\begin{figure} [h]
    \includegraphics [scale=0.6]{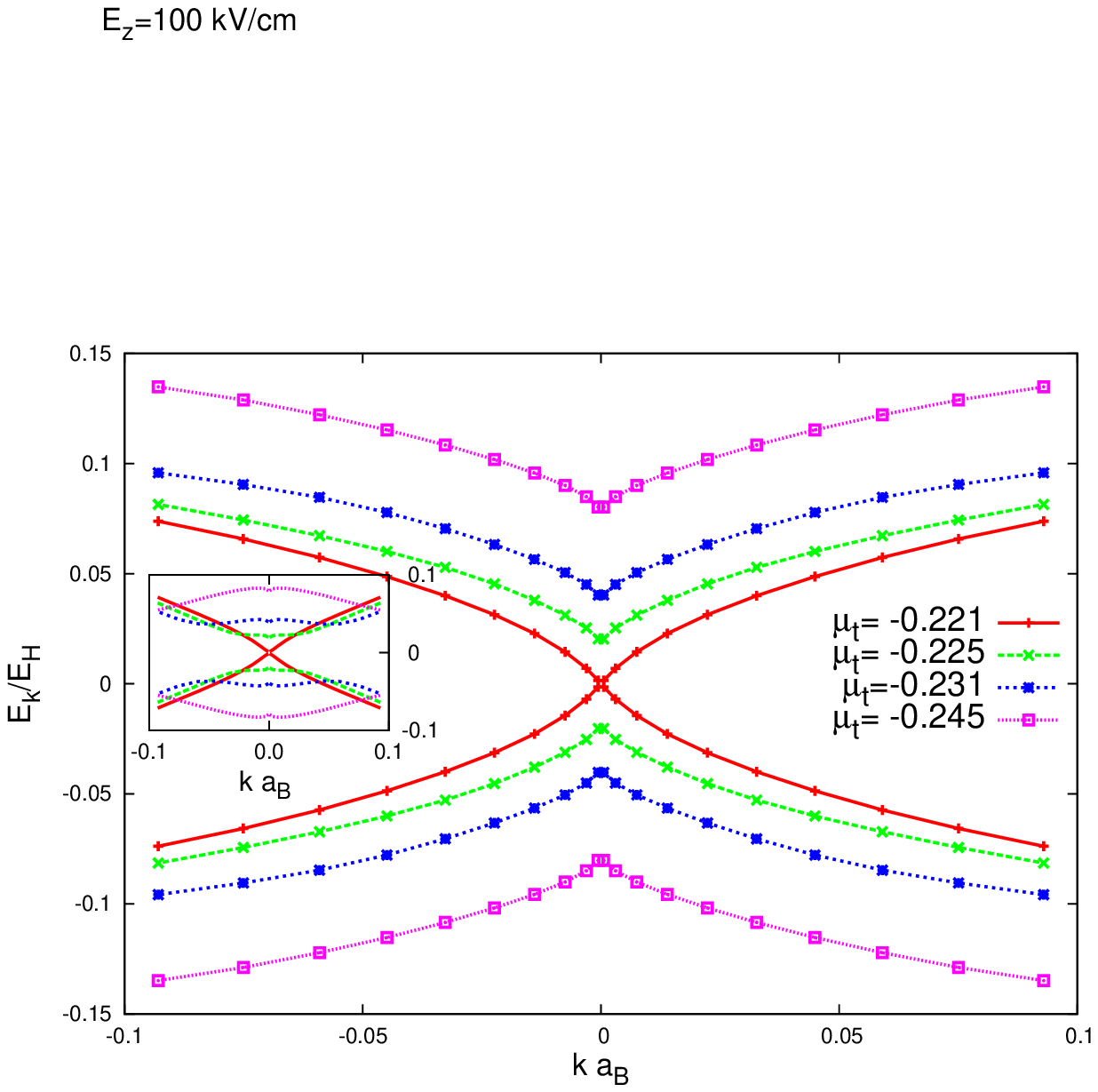} 
\label{figur}\caption{(Color online) The $E^{\pm}_k$ of the pure ESP triplet (${\it t}_{ESP}$) solution (case-iv in Table.I) around the Dirac point at $\mu_c=-0.221$ black-curve with $+$ signs. The main plot is $ E^{+}_{k} $ and the inset is $ E^{-}_{k}$. }
\label{triplet_evolution}
\end{figure}

A pure triplet $t_{ESP}$ superconductor, i.e. [case-iv in Table.(\ref{TRS_cases})], can be, in principle, obtained in the minimal model even for a finite SOC if the pairing potential has no s-channel, i.e. $V_s=0$ in Eq.\ref{singlet_triplet_int}. In Fig.\ref{DOS} the exact solution of the energy bands and the energy DOS for this case are shown where the data from the pure {\it s} (case-i and iii with $V_t=0$) and the mixed s-t phases (case ii with $V_s, V_t\ne 0$) are also shown for comparison. In the $t_{ESP}$ solutions, the SOC and the particle concentration $n_x$ are tuned in each case (a,d), (b,e) and (c,f) so that the critical Fermi level $\mu_c$ is at $k=0$ where a Dirac-like spectrum is observed. The Fig.\ref{triplet_evolution} is somewhat complementary to the picture presented in Fig.\ref{DOS} in that, the evolution to/from the Dirac-like spectrum of this $t_{ESP}$ superconductor is also shown in the vicinity of the critical Fermi level $\mu_c=-0.221$.

The topology on the other hand, is encoded in the nodes ofthe pair potential, i.e. $\tilde{\Delta}^{\pm}_k=\vert \psi_k \mp \gamma_k F_k\vert$.  
The number as well as the position of these nodes are determined by the full momentum dependence of the t/s ratio $\vert F_k/\psi_k \vert$. Depending on this ratio, there can be  zero, one or more point or line nodes in each branch of $\tilde{\Delta}^{\pm}_k$. The pairing symmetries together with the SOC and $\mu$ determine which of these nodes to appear in which branch of $E_k^\pm$.

Fig.\ref{mixed_OPs_bands}.(a,d,g,j) for $V_1(q)$ has RLNs in both branches of the gap $\tilde{\Delta}_k^{(\pm)}$ whereas no line nodes are observed for $V_2(q)$ in (b,e,h,k) due to the strong singlet. The RLNs can be observed for $V_3(q)$ as shown in (c,f,i,l) for $\tilde{\Delta}_k^{(\pm)}$ provided the singlet is weakened further by a repulsive hardcore interaction. Generally, RLNs shift to higher $k$ for increasing $\mu$ whereas they shift towards $k=0$ for larger SOC. We believe that the energy line nodes reported for $BiPd$ \cite{mixed3,BiPd} ,$ Y_{2}C_{3} $ \cite{mixed1}  and $ CePt_{3}Si $ \cite{PRL94_197002,PRL94_207002} may be RLNs.               

 In NCS, the connection between the nodes and the topology has been widely studied under strong anisotropy where nodes appear in specific angles in the $k$-space dictated by the tetragonal symmetry.\cite{NCS_nodal_topo0,NCS_nodal_topo,NCS_nodal_topo2,NCS_nodal_topo3} On the other hand, RLNs, favoring continuous rotational symmetry are complementary to these well studied examples. It is therefore expected that the RLN topology has properties distinctively different from those appearing in strongly anisotropic systems and this has not been studied yet. The RLNs can exist in a pairwise continuous and closed set of $k$-space points and can be simultaneously present in a multiple of radial locations in the $k$-space. Our analysis reveals that the number (even or odd) as well as the position of the RLNs in the pair potential with respect to the Fermi level is crucial in the determination of the band topology.

\begin{figure}
\includegraphics [scale=0.7]{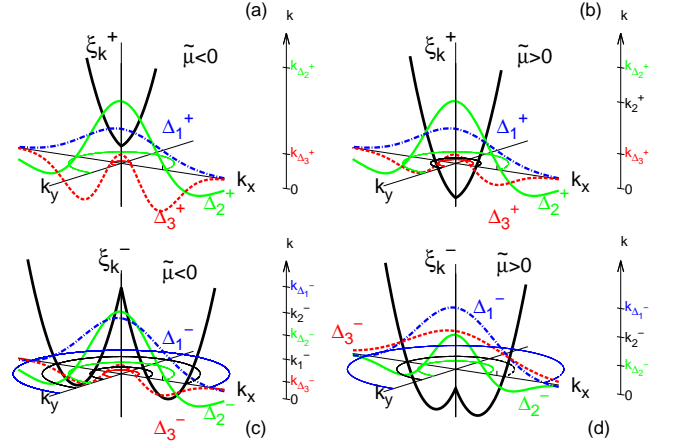}
\label{FP}\caption{(Color Online) Possible cases regarding the position of the Fermi momenta $k_i^\lambda$ of the relevant spin orbit branch $\lambda=\pm$ where $\tilde{\xi}_k^\lambda\vert_{k=k_i^\lambda}=0$ and the position of the RLNs in ${\tilde \Delta}_k^\pm=\vert \psi_k \mp\gamma_k F_k \vert$. The thick lines in black indicate the $\tilde{\xi}_{k}^{\lambda}$ for $\lambda=\pm$. The thick colored lines indicate three different nodal behaviour for ${\tilde \Delta}_k^\pm$ as depicted by ${\tilde \Delta}_1^\pm, {\tilde \Delta}_2^\pm, {\tilde \Delta}_3^\pm$. The radial $k$ axis on the right of each figure indicates the relative positions of the Fermi wavevectors and the gap RLNs.} 
\label{zeros_and_nodes}
\end{figure}

The Fermi energy $\mu$ and the SOC determine the number of bands crossing the Fermi level. Writing $\tilde{\xi}_{k}^{\lambda}=\hbar^2(\gamma_k k- k_1^\lambda)(\gamma_k k- k_2^\lambda)/(2m)$ with $\lambda=\pm$, the number of bands crossing the Fermi level is given by the number of radially admissible ($k \ge 0$) solutions of $\tilde{\xi}_{k}^{\pm}=0$. Note that for given $k_1^\pm, k_2^\pm$, $\tilde{\mu}=-\hbar^2 k_1^+ k_2^+ /(2m)=-\hbar^2 k_1^- k_2^- /(2m)$ and $\pm \alpha=-\hbar^2(k_1^\pm + k_2^\pm)/(2m)$. Here $\gamma_k$ is a k-dependent sign which depends on the specific model. We consider here $\gamma_k=1$ which can dramatically simplify the analysis without loosing generality. On the other hand, for a given $\tilde{\mu}$ and $\alpha$  
\begin{eqnarray}
k_i^\lambda=&(m/\hbar^2)\Bigl[-\lambda \alpha +(-1)^i \sqrt{\alpha^2+2\frac{\hbar^2}{m}\tilde{\mu}}\Bigr]~,\qquad i=(1,2) \label{energy_nodes} 
\end{eqnarray}
are the zeros where $\tilde{\xi}_{k}^{\lambda}=0$. The $\alpha<0$ case swaps between the two $\lambda$ branches. We therefore confine our analysis to $\alpha>0$ for simplicity. The Fermi wavevectors are the positive solutions in Eq.(\ref{energy_nodes}) which can be studied separately for $\tilde{\mu} >0$ and $\tilde{\mu} <0$. These are illustrated in Fig.\ref{zeros_and_nodes} (a,b,c,d) together with the nodal positions of ${\tilde \Delta}_k^\lambda$. For $\tilde{\mu}<0$ no Fermi wavevector is present in the $+$ branch (Fig.\ref{zeros_and_nodes}.a), whereas two Fermi wavevectors in the $-$ branch (Fig.\ref{zeros_and_nodes}.c) given by $(\lambda,i)=(-,1)$ and $(-,2)$. In the $\tilde{\mu} >0$ case, there is only one Fermi wavevector for each branch described by $(\lambda,i)=(+,2)$ (Fig.\ref{zeros_and_nodes}.b) and $(-,2)$ (Fig.\ref{zeros_and_nodes}.d). It is clear in Eq.(\ref{energy_nodes}) that the positions of $k_i^\lambda$ can be controlled externally by $\mu$ and $\alpha$. On the other hand, the pairing interaction is more effective on the k-dependent pair potential. The position $k_\Delta^\lambda$ i.e. ${\tilde \Delta}_k^\lambda|_{k=k_\Delta^\lambda}=0$ can only be obtained from the results of self consistent mean field Eq's(\ref{coupled_OPs}). Hence, the orientation of $k_i^\lambda$ relative to $k_\Delta^\lambda$ can be different for a given $\mu, \alpha$ and the pairing interaction. Distinct cases are depicted on the right side of each plot in Fig.(\ref{zeros_and_nodes}). 
  
\begin{figure}
\includegraphics [scale=0.52]{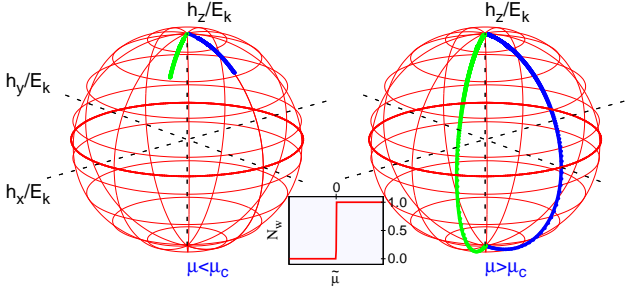}
\label{FP}\caption{(Color online) The trivial ($\mu <0$) and the nontrivial ($\mu>0$) topologies.The green/blue paths are followed by $\hat{\bf n}_{\bf k}$ as $k$ is brought from $\infty$ to zero at two different $\phi_{\bf k}$. The inset is $N_w(\mu)$ in Eq.(\ref{chern_index}).} 
\label{unit_sphere}
\end{figure}

The topological characterization of the energy bands has been thoroughly investigated previously in the presence of non-spatial symmetries\cite{topo_class_nonspat}, i.e. TRS, $\Lambda$ and the FX in the context of this article. According to the Altland-Zirnbauer classification, this corresponds to DIII class where the two dimensional ones are topologically characterized by the $Z_2$ indices. This picture was extended to include the discrete spatial symmetries such as reflections where an increasing number of references can be found\cite{topo_class_nonspat+spat}. 
The fully gapped ones can be characterized by global topological invariants. The fully gapped superconductors with a generic Hamiltonian ${\cal H}_{\bf k}=\sigma.{\bf h}_{\bf k}$ where the vector $\vert {\bf h}_{\bf k} \vert \ne 0$ for all $ \bf k$ points can be described by global topological invariants. The best known of all these is known as the Chern index\cite{Chern_index}   
 \begin{eqnarray}
N_{w1}=\frac{1}{8\pi}\int\,d^2k\,\epsilon_{ij}\,{\hat n}_{\bf k}.\Bigl(\frac{\partial{\hat n}_{\bf k}}{\partial k_i} \times \frac{\partial{\hat n}_{\bf k}}{\partial k_j} \Bigr)
\label{chern_index}
\end{eqnarray} 
describing the topological invariant in the two dimensional mapping ${\bf k} \to {\hat n}_{\bf k}={\bf h}_{\bf k}/|{\bf h}_{\bf k}|$. For instance, Eq.(\ref{chern_index}) can be applied to the  $^3$He-A phase \cite{Volovik} in which Hamiltonian is similar to Dirac-electron in 2+1 dimension i.e. ${\bf h}_{\bf k}=(\Delta_0 k_x,\Delta_0 k_y,\epsilon_k) $ with $ \epsilon_k=\hbar^2 k^2/(2m)-\mu $. Starting from the north pole ${\hat n}_{\infty}=(0,0,1)$ at $k \to \infty $, ${\hat n}_{\bf k}$ ends up in the $k \to 0$ limit either in the north pole when $\mu <0$, or it wraps the full sphere before ending up in the south pole when $\mu>0$ as shown in Fig.{\ref{unit_sphere}}.     

In the presence of RLNs the additional information about the position of $k_\Delta^\lambda \ne 0$ relative to the Fermi level is important in the topological characterization. In this regard, Fig.\ref{zeros_and_nodes} illustrates how this can be done. All distinct positions of $k_\Delta^\lambda$ relative to the Fermi level are  indicated for $\tilde{\mu}<0$ in (Fig.\ref{zeros_and_nodes}.a) and (Fig.\ref{zeros_and_nodes}.c) and $\tilde{\mu}>0$ in (Fig.\ref{zeros_and_nodes}.b) and (Fig.\ref{zeros_and_nodes}.d)  on the vertical $k$-axes in each figure. Three different gap profiles, indicated by $\Delta_j^\lambda$ for (j=1,2,3), are also shown in each plot. An equivalent form of Eq.(\ref{chern_index}) is integral over the solid angle $N_{w1}=\int\,d\Omega_{{\hat n}_{\bf k}}/(4\pi)=\int\,d^2k\,J(\theta,\phi)/(k_x,k_y)$ where $d\Omega_{{\hat n}_{\bf k}}=d(\cos\theta) d\phi$ with ${\hat n}_{\bf k}={\hat n}(\theta,\phi)$ and $J(\theta,\phi)/(k_x,k_y)$ is the Jacobian of the transformation ${\bf k} \to {\hat n}_{\bf k}$. In the presence of RLNs, the Eq.(\ref{chern_index}) is therefore not integer-valued if one considers the full ${\bf k}$-space. An integer index can be obtained however, if $k_\Delta^\lambda \le k < \infty$ is considered.

A second method was proposed in Ref.'s\cite{NCS_nodal_topo0,NCS_nodal_topo,NCS_nodal_topo2} for Hamiltonians respecting "chiral symmetry". The "chiral symmetry" $\chi$ is the product of the TRS and the particle hole symmetry. Since both symmetries are preserved in this work the chiral symmetry is also manifested. In these systems a new topological index can be defined by bringing the Hamiltonian in Eq.(\ref{hamilt_1}) into the off-diagonal form. This can be done by a unitary transformation $V$ as,\cite{NCS_nodal_topo0,NCS_nodal_topo,NCS_nodal_topo2}  
\begin{eqnarray}
V{\cal H}_{\bf k}V^\dagger=\pmatrix{0 & D_{\bf k} \cr 
  D_{\bf k}^\dagger & 0} ,\qquad   V=\frac{1}{\sqrt{2}}\pmatrix{\sigma_0 & -\sigma_2 \cr 
  i \sigma_2 & i \sigma_0} 
\label{off-diag1}
\end{eqnarray} 
where $D_{\bf k}=C_k[cos(\phi_k) \sigma_z+ i sin(\phi_k)\sigma_0]-i B_k \sigma_2$  with $C_k=|{\bf G_k}|-i F_k$ and $B_k=\xi_k+i \psi_k$. Here $D_{\bf k}$ is well defined only in those ${\bf k}$ points when the energy spectrum Eq.(\ref{e_values_m}) is nonvanishing. Hence this method applies also when the energy is fully gapped. For such NCS Hamiltonians as in Eq.(\ref{off-diag1}) a {\it momentum-dependent} topological index was defined in Ref.'s\cite{NCS_nodal_topo0,NCS_nodal_topo,NCS_nodal_topo2,NCS_nodal_topo3} as 
\begin{eqnarray}
N_{w2}(k_\perp)=\frac{1}{2\pi} \Im m\Bigl\{ \int_{-\infty}^\infty dk_\parallel \, \partial_{k_\parallel} \ln det({\tilde{D}_{\bf k}}) \Bigl\}
\label{off-diag3}
\end{eqnarray} 
where $k_\parallel$ and $k_\perp$ are cooordinates fully parametrizing the ${\bf k}$-plane. Note that Eq.(\ref{off-diag3}) can be completed into a loop integral in the $k_x-k_y$ plane closing at infinity in the upper or lower half plane. Here we transformed $ D_{\bf k}\rightarrow {\tilde{D}_{\bf k}}$ as $det({\tilde{D}_{\bf k}})=det(D_{\bf k})/|det(D_{\bf k})|$.
For instance, if $k_\parallel=k_x$ then, $N_{w2}$ becomes a $k_y$ dependent index. Such a momentum dependent index cannot be defined globally. In the context of this work, the $\phi$-independence of $det({\tilde{D}_{\bf k}})$ allows $k_\parallel$ to be taken along any radial axis, i.e. $k_\parallel=k$. The Eq.(\ref{off-diag3}) can then be turned into  
\begin{eqnarray}
N_{w2}=\frac{1}{\pi} \Im m\Bigl\{ \int_{0}^\infty dk \, \partial_{k} \ln det({\tilde{D}_{\bf k}}) \Bigl\}
\label{off-diag2}  
\end{eqnarray}
where $N_{w2}$ is independent of $\phi$, hence a global topological index. The Eq.(\ref{off-diag2}) can now be shown to be connected with $N_{w1}$ in Eq.(\ref{chern_index}). Using the definitions of $D_{\bf k}, C_k$ and $B_k$, we have $det({D_{\bf k})}=({\tilde \xi}_k^+ +i{\tilde \Delta}^+_k)({\tilde \xi}_k^- +i{\tilde \Delta}^-_k)$ where ${\tilde \xi}_k^\lambda$ and ${\tilde \Delta}_k^\lambda$ for $\lambda=\pm$ are defined in Eq's.(\ref{coupled_OPs}) and (\ref{e_values_m}). The Eq.(\ref{off-diag2}) is therefore
\begin{eqnarray}
N_{w2}=\frac{1}{\pi} \sum_{\lambda} \int_0^\infty dk \,\partial_k [arg({\tilde \xi}_k^\lambda +i{\tilde \Delta}^\lambda_k)]\nonumber \\ 
= \sum_{\lambda} \int_0^\infty \frac{d\theta^\lambda}{\pi}~.
\label{winding_mixed2}
\end{eqnarray}
This firstly confirmes that each branch is characterized by a separate topological index $N_{w2}^\lambda$. Eq.(\ref{winding_mixed2}) has been obtained before in a different context\cite{NCS_nodal_topo3}. The U(1) phases entering Eq.(\ref{winding_mixed2}) are the polar angles $\theta^\lambda=tan^{-1}{\tilde \Delta}^\lambda_k/{\tilde \xi}_k^\lambda$ of the Hamiltonian unit vector $n^\lambda(\theta,\phi)=({\tilde \Delta}_k^\lambda \cos\phi,{\tilde \Delta}_k^\lambda\sin\phi,{\tilde \xi}_k^\lambda)$ at a fixed $\phi^*$. Eq.(\ref{winding_mixed2}) is therefore identical with the winding of the polar angle on the unit circle at a fixed longitude of the unit sphere  ${\hat n}_{\bf k}$ in Eq.(\ref{chern_index}). Considering the $\phi$ invariance in this work, there is therefore a one-to-one correspondence between Eq.(\ref{chern_index}) and Eq.(\ref{winding_mixed2}) [hence Eq.(\ref{off-diag2})]. 

In the case of RLNs, Eq.(\ref{winding_mixed2}) is also noninteger valued as Eq.(\ref{chern_index}), if the entire ${\bf k}$ plane is considered. 
The resolution is to restrict the integral to the maximum range again in order to yield a full coverage on the unit circle defined by ${\hat n}(\theta,\phi^*)$. This corresponds to the same reduced range $k_\Delta^\lambda\le k < \infty$ for each $\lambda$ separately. The angle $\theta^\lambda$ changes by $\Delta\theta^\lambda_{j+1,j}=-\pi[sgn({\tilde \xi}_{j+1}^\lambda)-sgn({\tilde \xi}_{j}^\lambda)]/2$ between the $j+1$'st and the $j$'th  nodal positions of ${\tilde \Delta}_k^\lambda$. Here ${\tilde \xi}_{j}^\lambda$ is the value of ${\tilde \xi}_{k}^\lambda$ at the $j$'th radial node position of ${\tilde \Delta}_{k}^\lambda$. With these boundaries of the integral in Eq.(\ref{winding_mixed2}), an integer valued index is obtained as $N_{w2}=\sum_{\lambda,j} \Delta\theta^\lambda_{j+1,j}/\pi$. 

Alternative techniques were proposed to extract the integer part of Eq.(\ref{off-diag3}). An equivalent definition of $N_{w2}$ makes use of the positions of the multiple sectors of the Fermi surface as shown in Ref's\cite{NCS_nodal_topo0,NCS_nodal_topo3} and assumes that the pair potential is sufficiently weak near the Fermi surface. Linearly expanding $\tilde{\xi}_k^\lambda$ and $\tilde{\Delta}_k^\lambda$ around the $i$'th Fermi surface position $k_i^\lambda$, the Eq.(\ref{off-diag2}) can be further simplified without loosing its topological characterization into a similar form used in the Ref.'s\cite{NCS_nodal_topo0,NCS_nodal_topo,NCS_nodal_topo2,NCS_nodal_topo3}, 
\begin{eqnarray}
N_{w2}=-\frac{1}{2}\sum_{k_i} sign[\partial_k ({\tilde{\xi}_k^+} {\tilde{\xi}_k^-})|_{k=k_i}]\nonumber \\ 
 sign[({\tilde{\Delta}_k^+ }{\tilde{\xi}_k^-}+{\tilde{\Delta}_k^- }{\tilde{\xi}_k^+} )|_{k=k_i}]
\label{winding_mixed}
\end{eqnarray} 
where point(s) $ k_i^\lambda$ are the Fermi momenta given by $ \tilde{\xi}^\lambda_k|_{k_i^\lambda}=0 $. Eq.(\ref{winding_mixed}) is a single topological index given for both branches. This expression can again be written as a sum of separate branches. To see this, it is sufficient to observe that either $ {\tilde{\xi}}_{k_i}^-=0$  or $ {\tilde{\xi}}_{k_i}^+=0$ in the right hand side of Eq.(\ref{winding_mixed}) and the corresponding terms can hence be discarded from the sum. The resulting expression of $ N_{w2} $ then becomes a sum over the independent branches as
\begin{eqnarray}
N_{w2}=-\frac{1}{2}\sum_{\lambda}\,\sum_{\tilde{\xi}^\lambda_{k_i}=0} sign[\partial_k {\tilde{\xi}_k^\lambda} |_{k=k_i}] sign[{\tilde{\Delta}_k^\lambda }|_{k=k_i}]
\label{winding_mixed1}
\end{eqnarray}  
  
\begin{figure}
\includegraphics [scale=0.5]{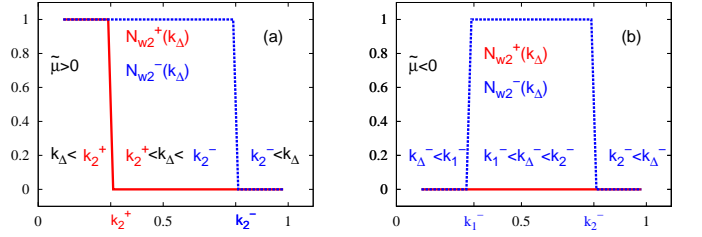}
\label{FP}\caption{(Color Online) The winding number $N_{w2}^\lambda$ in Eq.(\ref{winding_mixed1}) for a fixed ${\tilde \mu}$ when ${\tilde \mu}>0$ in (a) and ${\tilde \mu}<0$ in (b). The role of the relative position between the energy gap node $k_\Delta$ and the Fermi wavevector(s) in the determination of the topology in both branches is clearly observed. Here the simple notation $k_\Delta$ is used generically to mean $k_\Delta^\lambda$ when referring to a particular branch $\lambda$. For instance the cases $k_2^+<k_\Delta^+$ and $k_\Delta^-<k_2^-$ corresponding to $\tilde{\mu}>0$ can be summarized in the same plot by using the notation $k_\Delta$ only, as shown in (a).} 
\label{Z_2_indices_mixed}
\end{figure}

The Eq.(\ref{off-diag2}), equivalently Eq.'s (\ref{chern_index}) or (\ref{winding_mixed2}) in the reduced range, can produce all distinct topologies defined by the relative position of the gap node $k_\Delta^\lambda$ with respect to the Fermi wavevector $k_{1,2}^\lambda$ as shown in the vertical $k$-axes in Fig.(\ref{zeros_and_nodes}). We hence have $N_{w1}=N_{w2}$ in the reduced range. For example, the distinct cases given by $ \Delta_3^+ $ in  Fig.(\ref{zeros_and_nodes}b), $ \Delta_2^- $ in  Fig.(\ref{zeros_and_nodes}c) and $ \Delta_2^- $ in  Fig.(\ref{zeros_and_nodes}d) have non-trivial topology given by $N_{w2}=1 $ , whereas the other possibilities therein are trivial given by $ N_{w2} =0$. The topological indices corresponding to these possible configurations are summarized in Fig.\ref{Z_2_indices_mixed}. This figure can also demonstrate explicitly the independent topologies taken by different branches. For the ${\tilde \mu} >0$ case, the single Fermi wavevector $k_2^\lambda$ in each branch as indicated in Fig.\ref{zeros_and_nodes}.(c) and (d) can be in different positions on the k-axis. Depending on the position of $k_\Delta^\lambda$ winding number of each branch  are shown in Fig.\ref{Z_2_indices_mixed}.(a). On the other hand, if we assume $ k_2^\lambda>k_1^\lambda>0 $ a more interesting case occurs for the ${\tilde \mu} <0$ case where $+$ branch has no Fermi surface and hence has a trivial topology, whereas the $-$ branch has trivial topology for $k_2^- <k_\Delta^-$ and $k_\Delta^- < k_1^-$ and nontrivial topology for $k_1^- < k_\Delta^- < k_2^-$. Considering that the Fermi wavevector  position(s) relative to the energy gap node position(s) can, in principle, be controlled externally by ${\tilde \mu}$ and the SOC, our analysis here demonstrates that, the topological properties of the mixed NCSs are much richer than that in the pure triplet superconductor shown in Fig's \ref{triplet_evolution} and \ref{unit_sphere}. It should not be surprising that, controlling the topology, together with thermodynamic and other experiments sensitive to the energy density of states can be made in the near future in order to implement experimental as well as theoretical tools which can enhance our understanding the pairing potential(s) and the pairing mechanism(s).    
In summary, we investigated the most relevant unconventional pairing symmetries and the nodal structures in time reversal symmetric Hamiltonians with model pairing interactions and SOC under the general perspective of the $C_{\infty v}$ symmetry. Our results indicate that a strongly momentum dependent interaction (including the phonon originated ones) with a large attractive part in a TRS point ($q=0$ in the context of this work) can lead to a strong triplet pairing and the appearance of RLNs. Mixed, pure singlet and pure triplet solutions as well as their nodes at the level of the OP, the pair potential and the energy spectrum are investigated separately. In particular the nodal topology of the pure triplet superconductor between the trivial and the nontrivial cases can be manipulated by adjusting the Fermi level which can be experimentally accomplished by doping or by electrostatic gating. In a very recent work, such an external manipulation of the topology was shown experimentally for the topological $Z_2$ insulators\cite{Kouwenhoven}. On the other hand, the topological classification of the RLNs is shown to be noticeably richer than the other kinds of nodal superconductors which is an open field that can be further explored. With these at hand, we also put as a side remark that, EC with a strong SOC, which is one of the NCS models studied here, is a promising candidate in the near future where topological condensate in the mixed singlet-triplet state can be controllably accomplished in the context of Fig's.\ref{triplet_evolution}-\ref{Z_2_indices_mixed}.


\begin{thebibliography}{99}
\bibitem{basis} The basis will be understood as $\Psi_{\bf k}^\dagger=(\hat{e}_{{\bf
k} \uparrow}^\dagger~\hat{e}_{{\bf k} \downarrow}^\dagger~\hat{h}_{-{\bf k}
\uparrow}~\hat{h}_{-{\bf k} \downarrow})$ in the examination of the EC. It will be
assumed then that the electron-like, i.e. $\hat{e}_{{\bf k} \sigma},\hat{e}_{{\bf k}
\sigma}^\dagger$, and the hole-like, i.e. $\hat{h}_{{\bf k} \sigma},\hat{h}_{{\bf k}
\sigma}^\dagger$ states are related by the particle-hole symmetry.
\bibitem{EC_vs_NCS} The Rashba type SOC can be controlled by an external electric field. The exciton density can also be controlled by the driving laser creating the electron-hole pairs.
\end{thebibliography}
\end{document}